# Large Language Models in the UK: Public Use, Trust, and Attitudes

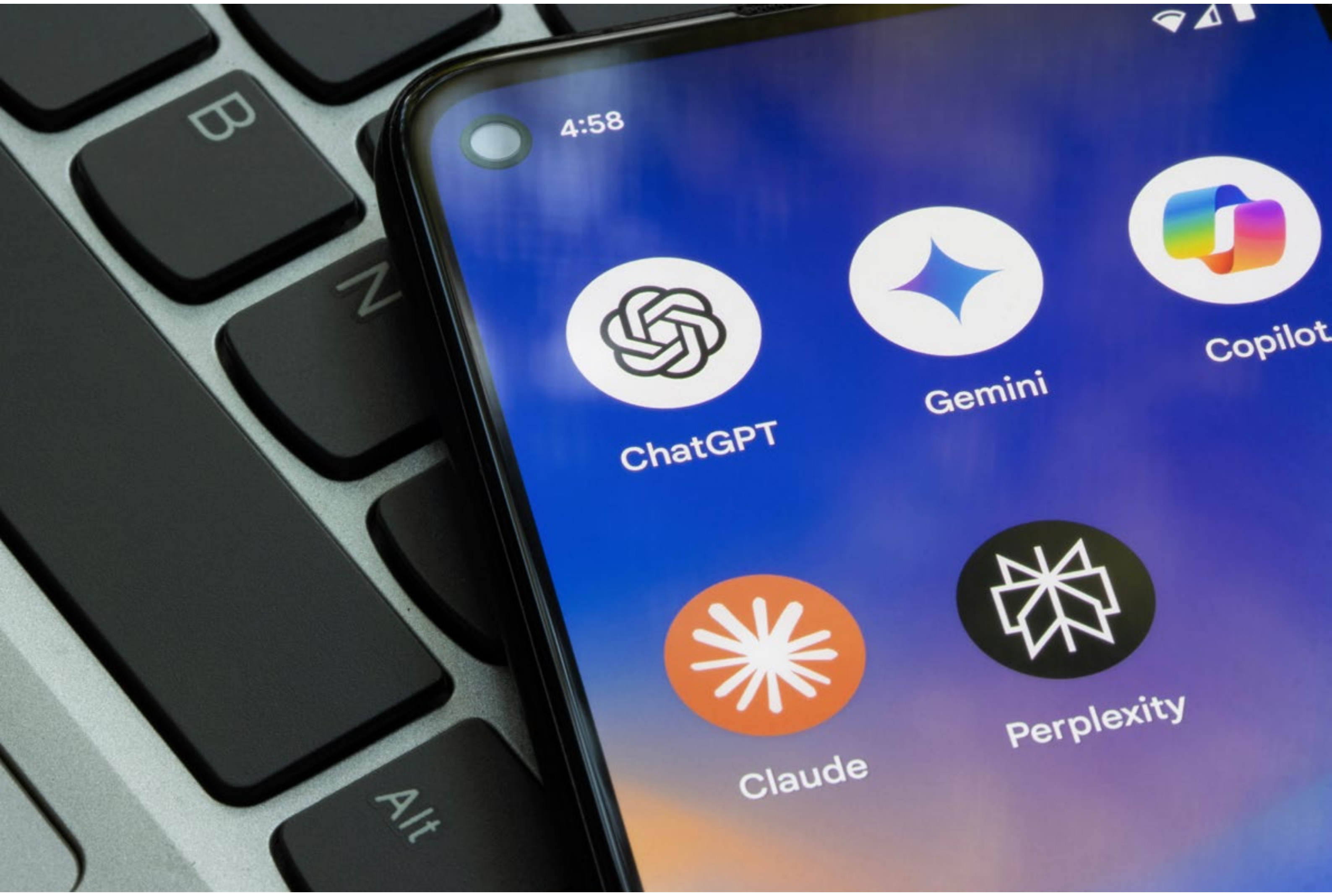

Florence E. Enock[1], Helen Z. Margetts[1,2]

**July 2026**

[1]Oxford Internet Institute, University of Oxford, Oxford, UK
[2]Data Science Institute, London School of Economics and Political Science, London, UK

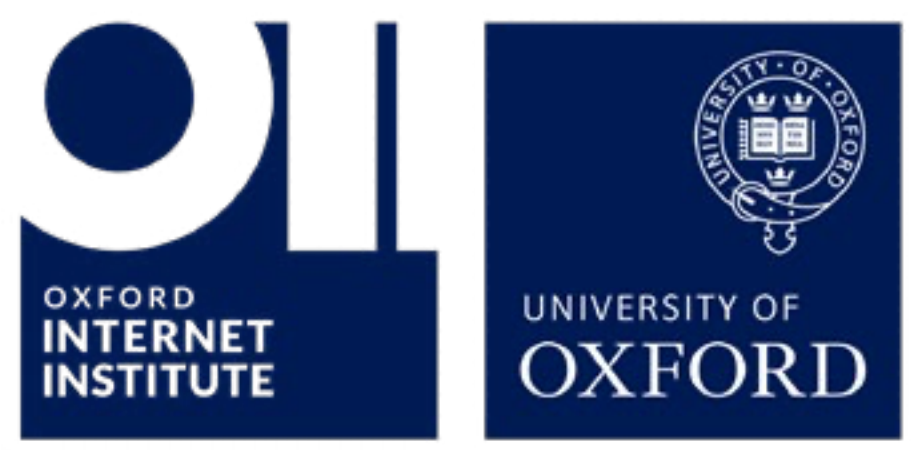

# Executive summary

Increasing numbers of people now routinely interact with large language models (LLMs) across many aspects of life, including in the workplace, educational settings, and for personal activities. The pace at which these tools have been adopted across society in recent years has led to substantial shifts in the ways in which people approach tasks and access information and advice. As a result, it is important for researchers and policymakers to gain up-to-date evidence on how the public engages with these technologies, including what people typically use them for, the extent to which users trust the information and advice that LLMs provide, and how the public perceives the potential benefits and risks associated with their use.

To explore these questions, we surveyed a nationally representative sample of 2,002 adults in the UK. Participants were asked about their use of LLMs, including both practical applications and more personal and social forms of engagement, their trust in the information provided by these systems across a range of topics and compared with other common sources, and their attitudes towards the potential societal benefits and risks associated with these tools. We also examined how patterns of use, trust, and public sentiment varied according to key demographic characteristics.

- **Our results show that while practical tasks remain the most common type of LLM use, many people now engage with these tools for emotional and social support.** Of respondents who report using LLMs at least weekly, a large majority say they use them for practical tasks such as generating text or factual searches very often (64%). However, a notable proportion report turning to LLMs for personal advice, support and companionship. Almost one third of regular users (31%) say they use LLMs for personal and emotional support, such as talking through problems and asking for help with decisions, while one quarter report interacting with LLMs for meaningful conversation at least occasionally. Younger adults are more likely to use LLMs across all five types of activity that we asked about, while higher educational attainment is associated with LLM use for practical, but not social, applications.
- **We also find that trust in LLM-generated information is relatively high.** When asked about the extent to which they typically trust information generated by LLMs, most respondents report at least some level of trust across a range of topics. Between 63% and 70% say they trust the accuracy of LLM-generated information at least somewhat when seeking information about public services, elections, climate change, health issues, and legal matters. While trust in LLMs is lower for advice relating to personal relationships, more than a third (38%) report trusting LLMs at least somewhat in this domain. Despite these high levels of trust, experts, online search, and friends and family remain more trusted sources of information across most topics, while LLMs are trusted more than social media and traditional news sources.
- **Finally, we show that public attitudes towards LLMs are characterised by both optimism and concern.** While over three quarters of our sample report feeling enthusiastic about the potential benefits of LLMs (77%), a majority also express concern about their potential risks (70%). Although respondents are more likely to believe that LLMs will create more benefits than harms than more harms than benefits, the findings on the whole suggest that people typically recognise both the opportunities and challenges associated with generative AI, rather than holding exclusively positive or negative views. Men are consistently more positive than women, showing that public sentiment towards LLMs is not uniform across social groups.

**KEYWORDS:**
Large Language Models, Generative AI, Human-AI Interaction, AI Adoption, Trust in AI, AI Attitudes

# Contents

# 1. Introduction

## 1.1 PUBLIC ADOPTION OF LLMS

Since the public launch of ChatGPT in late 2022, large language models (LLMs) have been rapidly adopted across societies for a wide range of uses. Advances in multimodal capabilities alongside the increased availability of publicly accessible models and mainstream integration into existing digital systems have further meant a sharp acceleration in the uptake of generative AI tools in recent years. By early 2026, it was estimated that 17.8% of the world's working-age population already used generative AI, with uptake particularly concentrated in middle- and high-income countries [1]. In the UK, the 2025 wave of the Ada-Turing public attitudes to AI survey found that 40% of the public reported having used an LLM before [2], while a 2026 study conducted by Ipsos for the BBC found this to have increased to 58%, with 35% reporting at least weekly use [3].

The widespread adoption of generative AI is seen across both work-related and personal activities. At the end of 2025, one quarter of US employees reported using AI frequently for work [4], while in the UK, the use of generative AI tools amongst public sector workers, including healthcare professionals, was already widespread by 2024 [5, 6]. A 2026 Ofcom study found that work, study and general information-seeking were the most commonly reported types of LLM activity [7]. As well as these practical applications, separate reports highlight the growing use of (and concerns around) LLMs for personal advice and emotional support [8].

Despite strong interest across academia, policy, and industry in the public adoption of generative AI [9, 10, 11], gaps remain in our knowledge about the specific ways in which people typically engage with these systems. In particular, there is growing interest in the potential psychosocial impacts of LLM use for social and emotional purposes [12, 13], yet few population-level estimates exist regarding the prevalence of these forms of engagement relative to practical use. Further, little is known about the social and demographic characteristics associated with different forms of LLM interaction. We address this gap by presenting survey data on overall LLM uptake, as well as the frequency with which users typically engage in a range of activities, from practical tasks to social companionship. We also examine key demographic characteristics associated with each type of use.

## 1.2 PUBLIC TRUST IN LLMS

Alongside the importance of understanding public uptake of generative AI, including who adopts these systems and for what purposes, it is also crucial to examine public trust in the information they provide. The widespread availability of generative AI tools has rapidly changed the ways in which people search for and access information online [14]. Already, a substantial proportion of users report turning to LLMs for information across a wide range of topics, from general everyday queries such as those relating to news or entertainment, to domains with potentially significant personal consequences, such as legal, financial, or health-related matters [15, 16]. Further, even individuals who do not actively seek out LLMs when searching for information are frequently exposed to AI-generated summaries through integration into existing search engines and other digital platforms [17].

While some evidence suggests that information seeking is now the lead use of generative AI across several countries [18], less is known about the extent to which people trust the accuracy of the information these systems provide. This is important because although LLMs can be highly useful in offering accessible and personalised information across many topics, they can also generate outputs that are biased, misleading, or inaccurate [19, 20]. Given that trust is likely to shape the extent to which people rely on and act upon given information, it is important to understand not only the extent of LLM use, but also how trust in the information provided compares with trust in other sources. To do this, we ask participants to rate their trust in the accuracy of information from a range of sources, including experts, online search, newspapers, social media, and LLMs, across multiple areas, including those relating to politics, science and health. This allows us to compare trust in LLM-provided information with trust in alternative sources and to examine how these patterns vary across topics. We also examine the demographic and social characteristics associated with trust in LLMs, which is of interest given their increasing use in domains where inaccurate or misleading information may carry adverse consequences.

## 1.3 PUBLIC ATTITUDES TOWARDS LLMS

Another key aim of this work is to provide an up-to-date overview of public sentiment towards generative AI. While several studies conducted across multiple countries have examined public attitudes towards AI (e.g., [2, 10, 21, 22, 23]), these have often focused broadly on AI and digital technologies, with few studies specific to public interaction with LLMs. Additionally, some of these studies were conducted before LLMs became widely accessible to the public, meaning that they capture views on different conceptions of AI (e.g., face recognition tools or targeted online advertising) and are unlikely to generalise to public experiences with LLMs. One exception is the second wave of the Ada–Turing Public Attitudes to AI survey [2], which examined public awareness, use, and perceptions of LLMs alongside a range of other AI applications and technologies. The findings suggested a broadly positive view of LLMs among the UK public, with 63% of respondents perceiving these tools as very or fairly beneficial, though with 47% expressing concern about their use. Work elsewhere suggests that while people are relatively positive about the potential of LLMs to save time and help with learning, they are also concerned about their potential role in replacing humans and spreading inaccurate information [3].

Given that public attitudes are likely to shape willingness to engage with and adopt generative AI tools, it is important to understand how people perceive these technologies and the benefits and risks associated with them. As the capabilities, availability, and visibility of such tools continue to expand, public sentiment may also change over time, creating a need for regularly updated evidence. Understanding which groups of people hold more positive or negative views of these technologies is also important. Prior research suggests that women are often more cautious in digital environments [24] and are more sceptical of AI more broadly [25], possibly because they are often disproportionately affected by technology-facilitated harms [26]. More negative perceptions of LLMs could contribute to lower engagement and, as a result, unequal access to the benefits these technologies may provide. We therefore examine current public perceptions of both the benefits and risks of LLMs, alongside the social and demographic characteristics associated with these attitudes.

## 1.4 RESEARCH CONTRIBUTION

This report presents findings from a nationally representative survey of 2,002 UK adults and, in doing so, makes four key contributions.

First, it provides evidence on how people currently engage with LLMs in practice, spanning both functional uses and more social forms of engagement. Given growing discussion around the use of generative AI to simulate human relationships and social connection, we compare five types of activity, ranging from practical tasks such as generating text and planning projects to social uses such as emotional support, companionship, and roleplay. We also explore perceptions of the humanness and gender of LLMs, given that anthropomorphism may contribute to greater social attachment [27].

Second, the work provides evidence on public trust in LLM-generated information. We compare trust in information provided by LLMs with trust in a range of other commonly used sources, including experts, search engines, newspapers, and social media, and examine how these patterns vary across different knowledge domains.

Third, the findings provide an up-to-date overview of public perceptions of the societal benefits and risks associated with LLMs, as well as overall sentiment towards these technologies. As generative AI tools become increasingly integrated into everyday life, gaining a current overview of public sentiment is crucial for anticipating patterns of adoption and understanding sources of public concern.

Finally, the report examines the social and demographic characteristics associated with LLM use, trust, and sentiment. This is important for understanding who is most likely to adopt and benefit from these technologies, as well as which groups may be more hesitant to engage with them or more vulnerable to potentially adverse effects. Such differences have important implications for working to ensure that the benefits of these technologies are accessible across society.

The context of this research, public engagement with LLMs, is an area which is still new and evolving, particularly as more people begin using LLMs and these technologies become increasingly integrated into workplaces and existing digital systems. While the findings presented here provide a current overview, the patterns they describe are also likely to change over time. As such, they provide a valuable baseline against which future changes in use, trust, and public sentiment can be assessed.

# 2. Research Methodology

## 2.1 DATA COLLECTION AND AVAILABILITY

Data collection took place online during December 2025. The survey was created and administered using Qualtrics (www.qualtrics.com) and participants were recruited through Prolific (www.prolific.com). Informed consent was obtained at the start of the survey according to approved ethical procedures. The full survey and anonymised data are available on the OSF page for this work (osf.io/agyfb).

## 2.2 SAMPLE CHARACTERISTICS

A total of 2,002 participants completed the survey and passed standard checks for data quality. The sample was designed to be nationally representative of the population of the United Kingdom across age, gender and region using Prolific's representative sample tool. Respondents were between 19 and 93 years old, with a mean age of 46.6 (SD = 15.8). 1,024 participants identified as female (51.1%), 961 as male (48.0%), and 4 as non-binary, with the remaining 13 indicating 'Prefer to self-describe' or 'Prefer not to say'. 82.9% of our sample identified as White, 6.8% as Asian/Asian British, 6.1% as Black, African, Caribbean or Black British, 2.7% as Mixed or Multiple ethnic groups, with the remainder indicating 'Other' or 'Prefer not to say'[1]. 72.5% of our sample were employed, 18.8% were not in paid work (i.e., homemaker or long-term sickness) or retired, 3.9% were unemployed (actively seeking work), and 4% were students. 58.9% of our sample had a university degree or higher and respondents generally reported high levels of digital literacy, with 89% rating their digital skills as good or very good.

## 2.3 SURVEY

### 2.3.1 DEMOGRAPHICS AND BACKGROUND QUESTIONS

For each participant, we collected standard demographic information including age, gender, ethnicity, education level, employment status, political orientation, and digital literacy.

Age was entered as a numerical value with a minimum of 18 years. For

1 While participants indicated more specific ethnic identities on the survey, we combine them into broader categories to simplify reporting here.

gender, ethnicity, employment status, and education level, participants were asked to select the option that best described them from a list of standard categories. To measure self-perceptions of digital skills, participants rated their ability to use digital technologies on a four-point scale ranging from Poor to Very good. Political orientation was measured on a continuous 0–100 unmarked scale, from *Extremely left* to *Extremely right*, with *Centre* at the midpoint.

### 2.3.2 LLM USE

Throughout the survey, LLMs were referred to as AI chatbots, but we first provided a description of these along with examples (see Supplementary Information section A.1 for definitions).

Participants were first asked whether they had ever used an AI chatbot (*Yes, No, Unsure*), whether they had ever interacted with an AI chatbot in the form of an AI companion (*Yes, No, Unsure*), how often they typically use AI chatbots in general (6-point scale: *Only once or twice before* to *Daily*), how often they use AI companions (7-point scale: *Never* to *Daily*), and how long they have been using AI chatbots for (4-point scale: *Less than a month* to *More than a year*).

### 2.3.3 TYPES OF LLM INTERACTION

To measure types of LLM interaction, respondents were asked about the extent to which they use AI chatbots for the following five purposes (4-point scale: *Never* to *Very often*):

- As **practical tools** (e.g., for generating text, coding and other work tasks, or factual searches);
- For **collaboration and planning** (e.g., brainstorming, planning activities like home projects and meals, asking for TV or travel recommendations);
- For **personal and emotional support** (e.g., talking through relationship problems, or asking for help with important decisions);
- To hold **deep and meaningful conversations** (e.g., exchanging thoughts about religion and morality, or discussing topics of interest such as politics or music);
- For **role-play and character-based interactions** (e.g., chatting with AI companions that simulate particular roles and personalities, or to act out fictional scenarios).

### 2.3.4 MODES OF LLM USE

Respondents were asked whether they primarily use text or voice replies (*Mainly text*, *Mainly voice*, *Both equally*) and, where applicable, which type of voice they typically use (*Male*, *Female*, *Both*, *Gender neutral*, *Not applicable*). They were also asked whether they typically adjust the tone settings when they interact with AI chatbots (e.g., to make them more formal, friendly, or cynical - *Yes*, *No*, *Unsure*).

#### 2.3.5 GENDER AND HUMANNESS PERCEPTIONS OF LLMS

Respondents indicated how they perceive the gender of the chatbots they typically interact with (*More male than female*, *More female than male*, *Both male and female*, *Ungendered*, *Unsure*) and how human-like their interactions with LLMs typically feel (4-point scale: *Not at all* to *Completely*).

#### 2.3.6 TRUST IN LLM-GENERATED INFORMATION

Participants rated their level of trust in six information sources across six domains (4-point scale: *Not at all* to *Very much*). The information sources were: AI chatbots (LLMs), Experts (a domain-specific expert for each topic), Friends and family, Google / Other search, Newspapers and television, and Social media. The information domains were: Climate change, Elections, Health issues, Legal matters, Personal relationships, and Public services.

#### 2.3.7 ATTITUDES TO LLMS

Respondents rated their agreement with the statements: 'I feel enthusiastic about the possible benefits of AI chatbots' and 'I feel worried about the potential risks of AI chatbots' (4-point scale: *Strongly disagree* to *Strongly agree*). Overall attitudes towards LLMs were measured using a 0–100 continuous unmarked slider, anchored to the left at *LLMs will create more harms than benefits* and to the right at *LLMs will create more benefits than harms*.

### 2.4 SURVEY PROCEDURE

After providing informed consent, participants were taken through the survey questions. Only those who indicated that they had previously used an LLM were asked whether they had ever used an AI companion. Only participants who reported having used an LLM at least more than once or twice before were asked about frequency and duration of LLM use, the extent to which they use LLMs for different types of interaction, along with questions about settings and gender and humanness perceptions. All participants, regardless of prior use, answered questions about trust and attitudes toward LLMs. For the trust measures, participants were randomly assigned to respond to two of the six information domains.

### 2.5 ANALYTICAL APPROACH

We report descriptive statistics to summarise patterns of LLM use across the sample. Descriptive results are presented as proportions or summary statistics. To examine demographic predictors of LLM use, trust and sentiment, we ran a series of regression models. When examining predictors of each type of LLM use, we restricted analyses to regular users (defined here as those using LLMs at least once per week). For models examining trust in LLMs and overall attitudes, the full sample was included.

Details of variable coding, including the construction of binary variables and the scaling of continuous predictors, are described in the relevant sections of the Results.

# 3. Results

## 3.1 USE OF LLMS

92.7% of respondents reported having used an LLM at least once before. Of those who had previously used an LLM a majority reported using them frequently, with 25.7% using daily, 32.9% several times a week and 14.8% about once a week. Infrequent use is less common — 12.8% reported using LLMs less than once a week, 6.2% less than once a month, and 7.7% said they had only used an LLM once or twice before.

Of regular LLM users (defined as respondents using LLMs at least once a week), the majority reported that they have been using them for a substantial length of time. 55.5% reported that they started using LLMs more than a year ago, while 27.3% said they had been using them for about 6-12 months. More recent uptake is less common — 15.7% have been using LLMs for 1-6 months, and just 1.5% reported that they only recently started using them.

Taken together, we find that most people using LLMs do so at least several times a week, and most regular users have already been using these tools for more than a year.

## 3.2 USE OF AI COMPANIONS

20.2% of the whole sample reported having used an AI companion (for example, on character-based platforms such as Replika or Character AI) at least once before.

**Of respondents who had used an AI companion platform before**, 31.1% said that they had only used these platforms once or twice before. However, a substantial proportion reported using them frequently — 15.6% said they used these about once a week, 20.8% several times a week, and 7.6% daily. Figure 1 shows overall proportions of respondents in our sample using LLMs and AI companions (panel A), along with frequency of use for each (panels B and C).

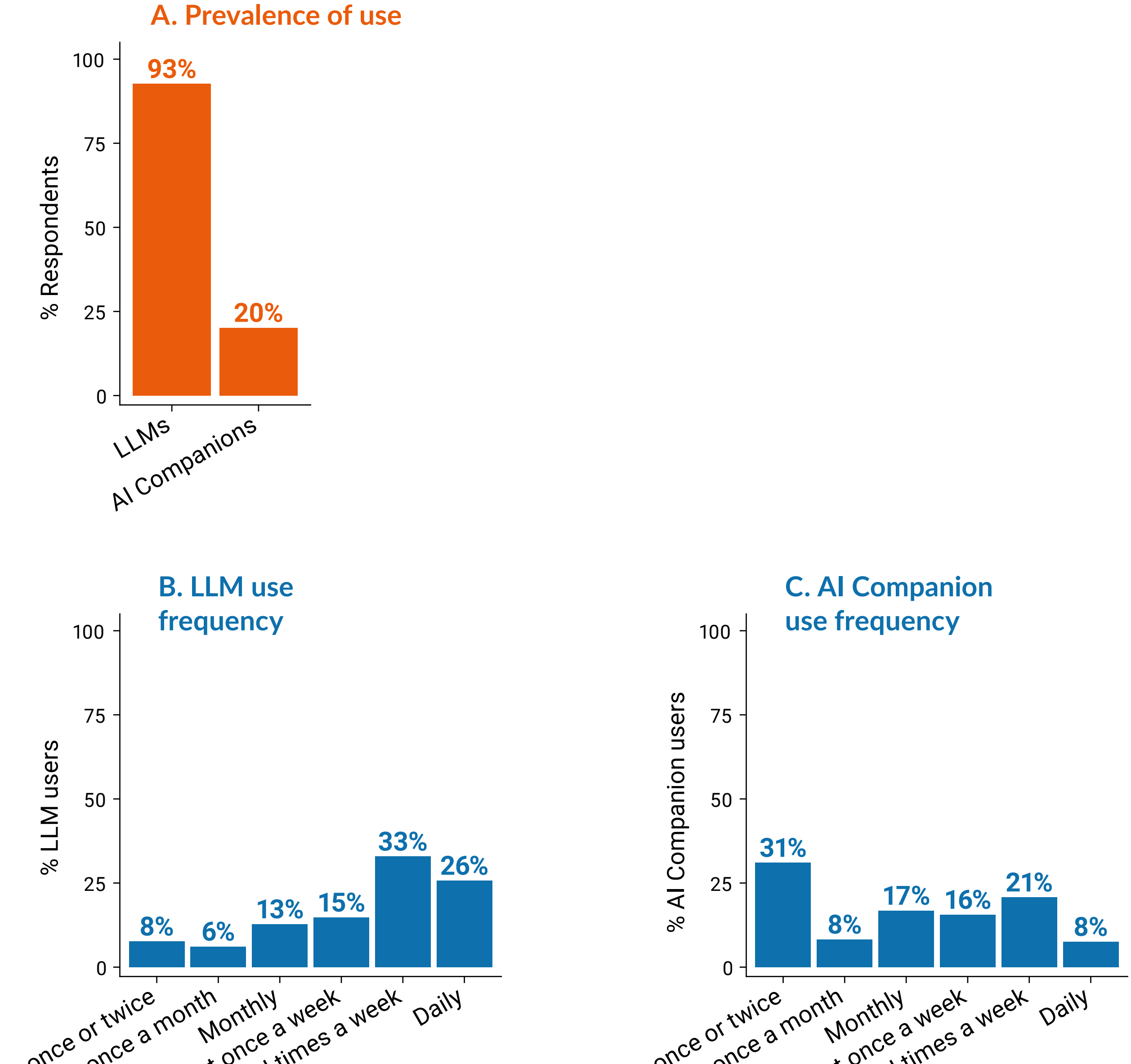

**Figure 1: Prevalence and frequency of use of LLMs and AI companions.** Panel A shows the proportion of all respondents reporting any prior use. Panels B and C show the frequency of use among respondents who reported at least some prior use of LLMs or AI companions.

## 3.3 MODES OF USE

**Of regular LLM users** (respondents using them at least once a week), a large majority reported doing so mainly using text replies (86.6%). Just 1% reported using mainly voice replies, and 11.8% reported using voice and text replies fairly equally. When using voice replies, respondents most commonly indicated using a female voice (34.4%), though with 26.9% using a male voice, 28.3% using both male and female voices, and 7.5% using a gender neutral voice.

When asked whether they typically change the tone settings of a chatbot's responses (for example to sound 'friendly' or 'cynical'), the majority of regular LLM users in our sample reported that they do not change the default settings (74.8%).

## 3.4 GENDER AND HUMANNESS PERCEPTIONS OF LLMS

When asked how human-like interactions with LLMs typically feel, regular users of LLMs (those using at least weekly) most commonly selected either 'slightly human-like' (44.9%) or 'moderately human-like' (36.2%). Just 12.1% said 'not at all human-like' and 5.3% said 'completely human-like'. Regular users most commonly reported perceiving LLMs as ungendered (46.6%), with 25.1% reporting that LLMs seem more male than female, and 15.7% as more female than male.

## 3.5 TYPES OF LLM USE

In the following section, we report the extent to which regular users of LLMs typically engage in the five different types of interaction that we asked about (see Methods, Section 2).

Regular LLM users most commonly reported using LLMs for practical tasks (such as for generating text, work-related tasks, or factual searches), with 64.1% using LLMs in this way very often and 27.2% using in this way occasionally.

Using LLMs for collaboration and planning (such as for brainstorming, planning activities and asking for recommendations) was also common, with 36.2% using LLMs in this way very often and 38.9% using in this way occasionally.

Interacting with LLMs more socially, for example for personal and emotional support, deeper conversation, or role-play and companionship was less typical, however, numbers of users engaging in these ways are still noteworthy. 31% of regular users reported interacting with LLMs for personal advice either very often or occasionally, 24.9% for deep and meaningful conversation either very often or occasionally, and 14% for role-play and companionship either very often or occasionally. Figure 2 (panel A) shows the extent to which regular LLM users report each of the five types of interaction.

### 3.5.1 DEMOGRAPHIC PREDICTORS OF TYPES OF LLM USE

We examined whether age, gender, education level and political orientation predicted each type of LLM interaction among regular users using five logistic regression models. For each model, the dependent variable was a dichotomous indicator of whether users reported a given type of LLM use at least occasionally (1 = Very often or Occasionally; 0 = Very rarely or Never). Prefer not to say responses were excluded.

Age and political orientation were included as continuous predictors and mean-centred, with coefficients rescaled to represent the effect of a 10-year increase in age and a 10-point increase on the political orientation scale. Gender was included as a binary predictor (1 = male, 0 = non-male), as was education level (1 = holds a university degree, 0 = does not hold a degree).

**Practical tasks:** Age and education level were significantly associated with using LLMs for practical tasks, with older adults less likely to engage in this way (OR = 0.78, $p < .001$) and those holding a degree more likely to engage in this way (OR = 1.86, $p = .003$). While there was a marginal effect of gender, with men slightly more likely to report this type of use (OR = 1.48, $p = .060$), there was no significant effect of political orientation ($p = .868$).

**Collaboration and planning:** Age and education level were again significantly associated with LLM use for planning and collaboration, with older adults less likely to report this type of use (OR = 0.82, $p < .001$), and those holding a university degree more likely to report this type of use (OR = 1.45, $p = .006$). There was no significant effect of gender ($p = .490$) or political orientation ($p = .094$).

**Personal and emotional support:** Age and gender were significantly associated with using LLMs for personal and emotional support, with older adults (OR = 0.74, $p < .001$) and men (OR = 0.68, $p < .002$) less likely to use LLMs in this way. There was no significant effect of education level ($p = .460$) or political orientation ($p = .726$).

**Deep and meaningful conversation:** Only age was a significant predictor of engaging with LLMs for deeper conversation, with younger respondents significantly more likely to report this type of use (OR = 0.84, $p < .001$). There were no significant effects of gender ($p = .361$), education level ($p = .386$), or political orientation ($p = .160$).

**Role-play and character-based interactions:** Finally, age, gender and political orientation were significantly associated with using LLMs for role-play and character-based interactions. Older respondents were less likely to report this type of use (OR = 0.77, $p < .001$), while men (OR = 1.51, $p = .014$) and more right-wing users (OR = 1.09, $p = .027$) were more likely to report this type of use (though with the magnitude of the latter effect extremely small). There was no significant effect of education level ($p = .119$). Figure 2 (panel B) shows significant demographic predictors of use across the five types of LLM interaction.

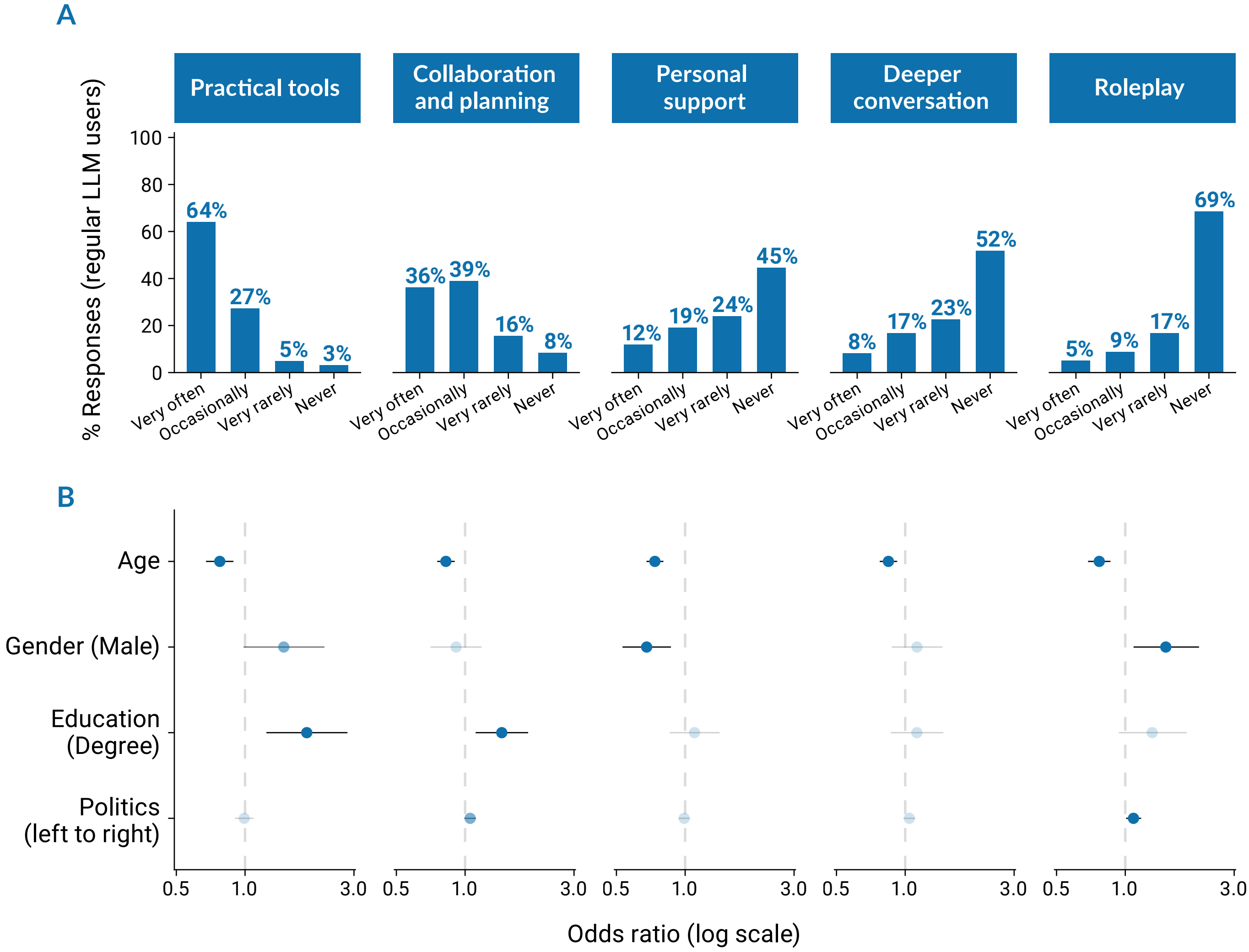

**Figure 2:** **Types of LLM use reported.** **Panel A shows the extent to which regular users of LLMs engage in different types of interaction. Panel B shows demographic predictors of each type of interaction from five logistic regressions. The points represent odds ratios with 95% confidence intervals, with opaque points indicating significant effects (p < .05), semi-transparent points indicating marginal effects (p <.10), and faded points indicating non-significant effects. The vertical dashed line indicates no effect. 'Prefer not to say' responses are excluded (all <1%).**

Taken together, age emerged as the most consistent predictor of engagement with LLMs across all interaction types. Odds ratios ranged from 0.74 to 0.84, indicating that with each additional decade of age, respondents had approximately 16–26% lower odds of using. Education level was associated with LLM use for practical tasks and planning, with those holding a degree showing 86% higher odds of using LLMs for practical tasks and 45% higher odds of using LLMs for collaboration and planning. Gender effects were less consistent: male participants had 48% higher odds of using LLMs for practical tasks and 51% higher odds of engaging in role-play and character-based interactions, but about 32% lower odds of using LLMs for personal support. Full results from our logistic regression models are shown in Table 1 in the Supplementary Information.

## 3.6 TRUST IN LLM-GENERATED INFORMATION

In the following section, we return to our whole sample to understand public trust in information given by LLMs across a variety of topics and compared to several sources.

Overall, trust in information given by LLMs was relatively high. For information about elections, public services, climate change, legal matters and health matters, the majority of respondents reported trusting LLMs at least somewhat. For information about elections, 63.0% of respondents indicated that they trust LLMs at least somewhat (15.6% very much), while for information about public services, 70.8% of respondents indicated that they trust LLMs at least somewhat (23.5% very much) and for information about climate change, 69.8% of respondents indicated that they trust LLMs at least somewhat (24.4% very much). For legal matters, 69.0% of respondents indicated that they trust LLMs at least somewhat (18.7% very much), while for health matters, 66.9% of respondents indicated that they trust LLMs at least somewhat (18.0% very much). Trust in LLMs was consistently higher than for social media and newspapers and television, but lower than for online search, experts, and friends and family (with the exception of the topic of climate change, where trust in LLMs was higher than in friends).

Trust in LLMs for advice on personal relationships was lower than for the other topics. However, 37.6% of respondents reported trusting LLMs at least somewhat (9.7% very much), which was again higher than for social media and newspapers and television but lower than for friends, online search and experts. Figure 3 shows the overall proportions of respondents trusting information provided by each source across the different domains.

### 3.6.1 SOCIAL AND DEMOGRAPHIC PREDICTORS OF TRUST IN LLMS

We ran a series of logistic regressions to examine the demographic predictors of trust in information provided by LLMs. We tested whether LLM use, age, gender, education level, and political orientation predicted trust in LLMs across six information domains. LLM use was coded as regular for respondents reporting at least weekly use, and the remaining predictor variables were coded as described above in Section 3.5.1.

LLM use was by far the strongest predictor of trust across all domains, with regular users between 3.27 and 4.54 times more likely to report trust in information given by LLMs than non-regular users (all $ps < .001$). Some additional patterns emerged, though with smaller effects. Older adults were more likely to report trust in LLMs for information about elections, legal matters, and health issues, with each additional decade of age associated with 12–14% higher odds of reporting trust ($ps < .05$). Political orientation showed small effects, with politically more right-leaning respondents around 11–12% more likely to report trust in LLMs for information about elections, legal matters, and personal relationships ($ps < .05$). Full results from the regressions are presented in Table 2 in the Supplementary Information.

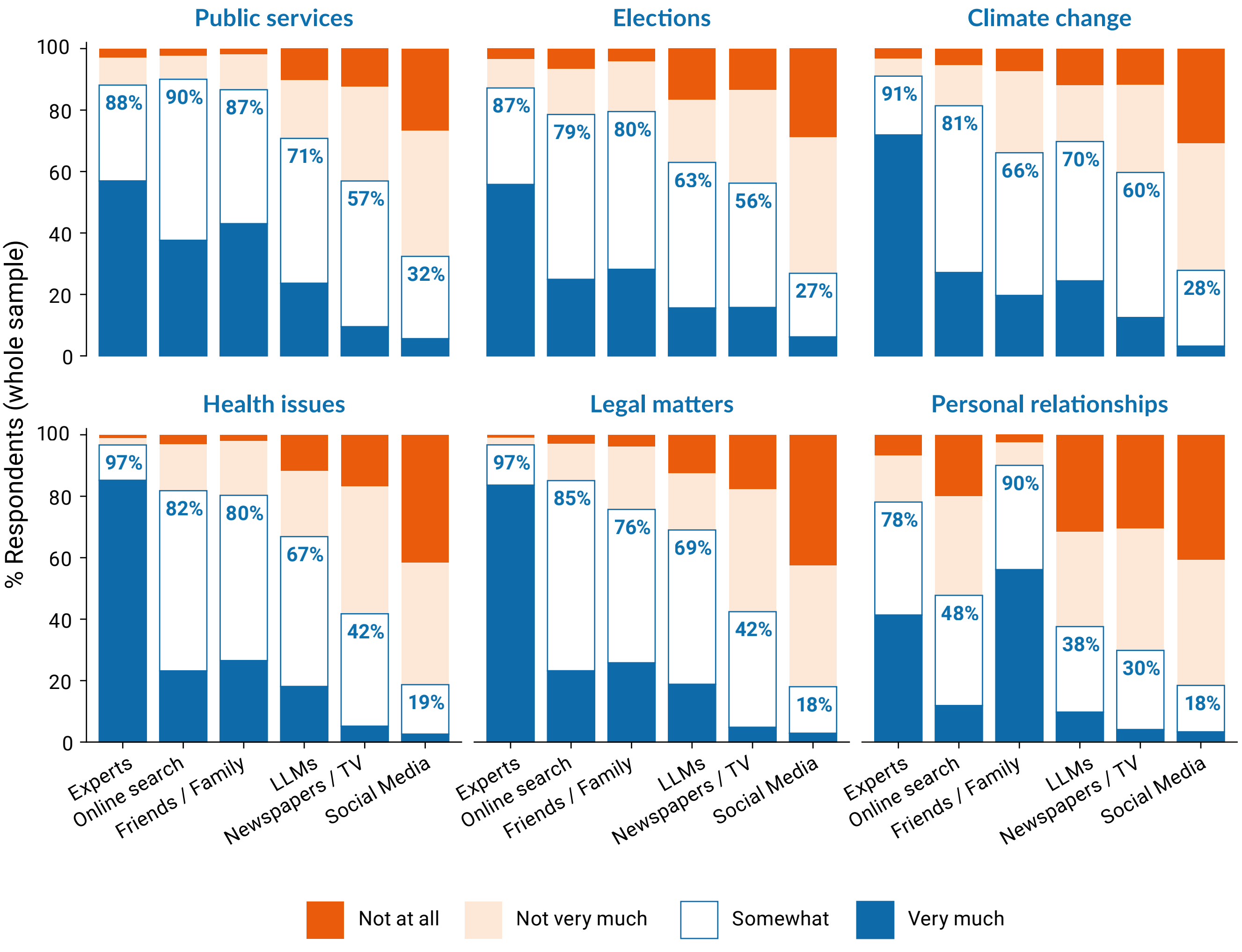

**Figure 3: Trust in LLMs relative to other sources by information topic.** **Bars show the distribution of trust ratings for each source (x-axis) within each domain (panels), with percentages in text showing proportions of respondents reporting at least some trust for each ('Somewhat' and 'Very much' combined). A majority of respondents reported trusting information given by LLMs across most topics, with the exception of advice about personal relationships. 'Prefer not to say' responses are excluded (< 1%).**

## 3.7 ATTITUDES TO LLMS

On the whole, respondents in our sample held fairly positive attitudes to LLMs. A large majority agreed that they 'feel enthusiastic about the possible benefits of AI chatbots' (35.3% strongly agree, while 42.0% agree somewhat). However, these positive attitudes were coupled with caution — a majority also agreed that they 'feel worried about the potential risks of AI chatbots' (30.8% strongly agree, and 38.8% somewhat agree).

When asked to rate their position on a continuous slider (0-100), with 'AI will create more harms than benefits' at the lower end and 'AI will create more benefits than harms' at the upper end, a majority of responses clustered above the midpoint of the scale (mean = 56.7 and median = 60), showing overall attitudes to LLMs were generally slightly more positive than negative, though with considerable spread. Figure 4 (panel A) shows the distribution of these attitude scores.

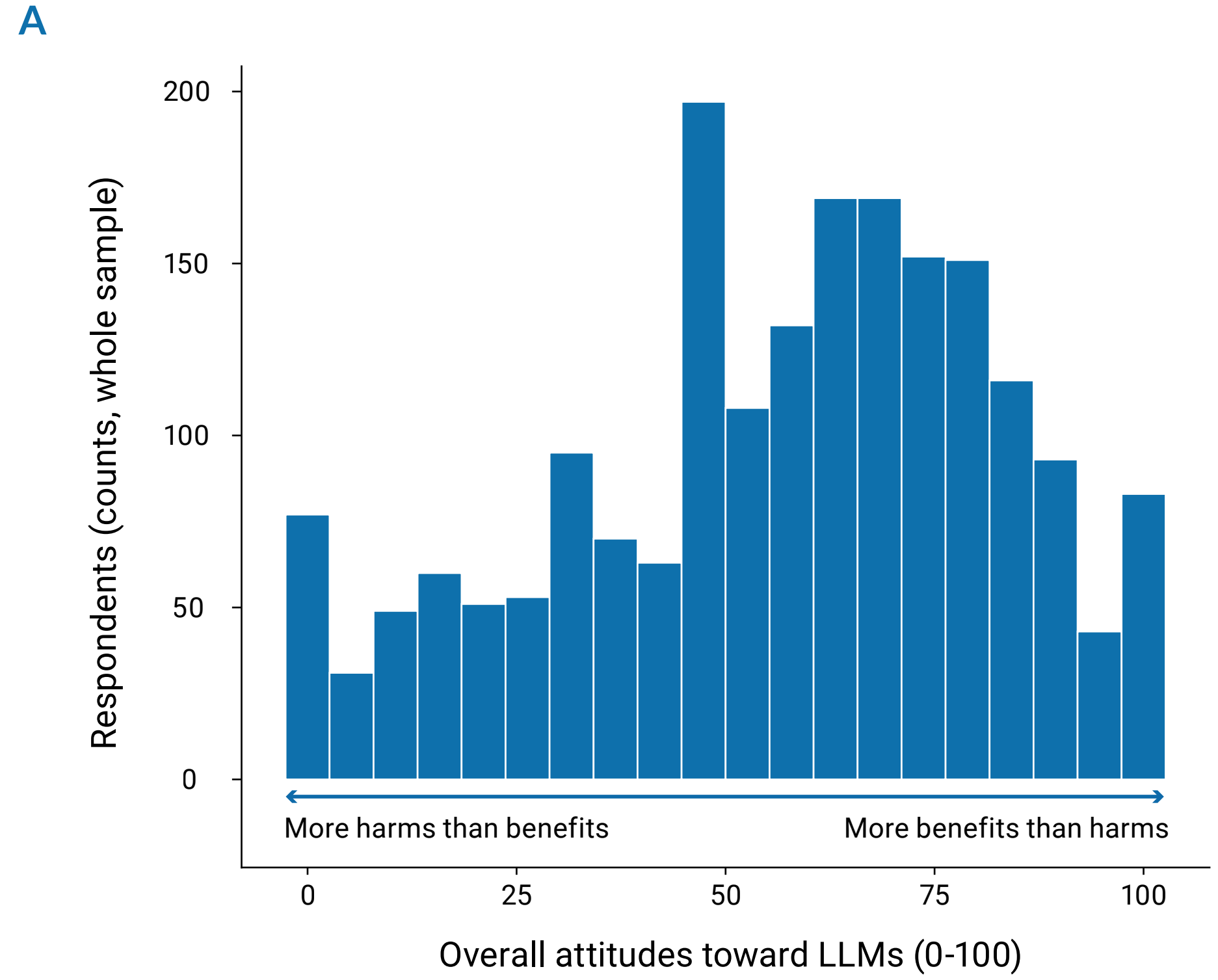

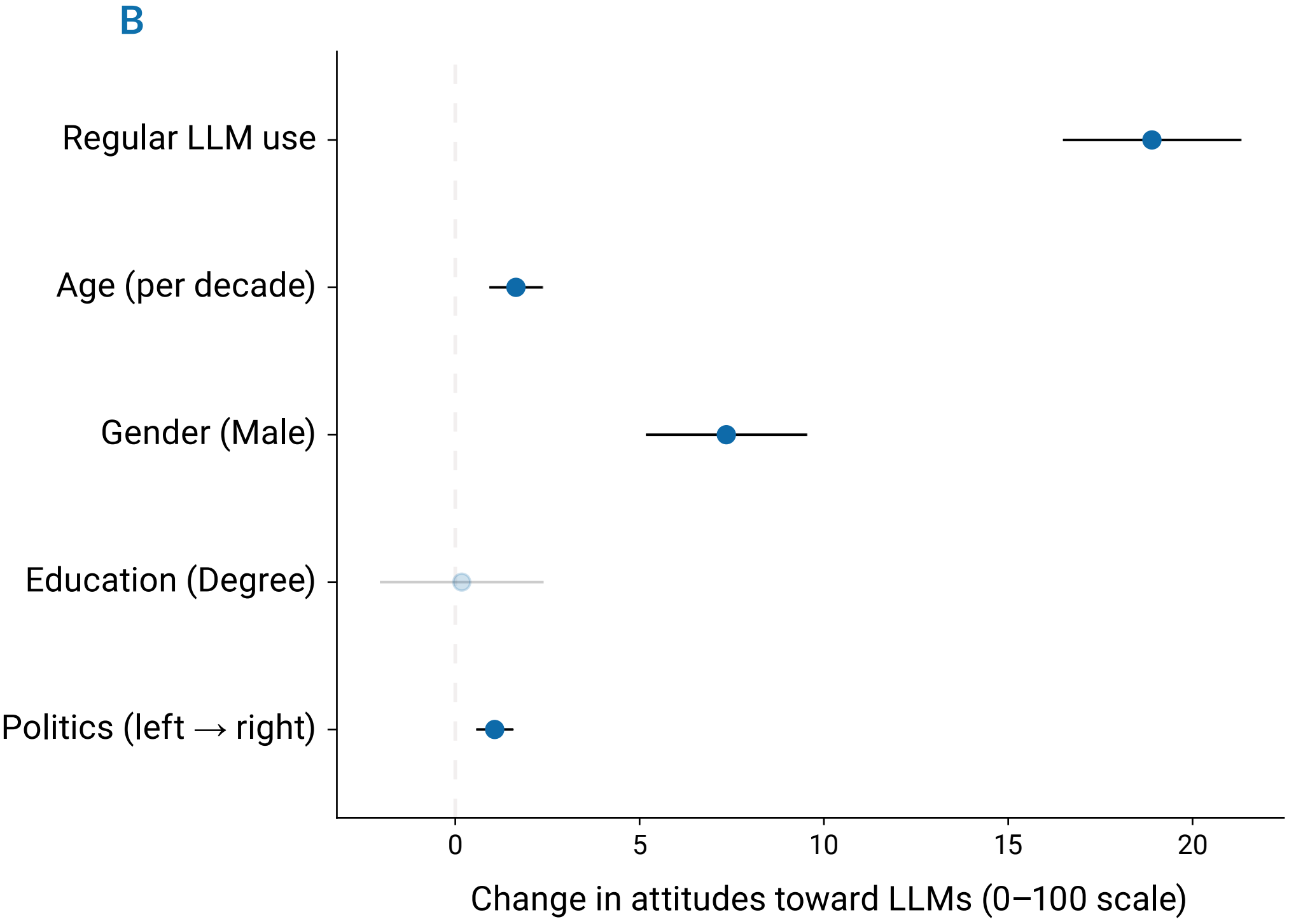

**Figure 4: Attitudes toward LLMs and corresponding demographic predictors.** Panel A shows the distribution of overall attitudes toward LLMs across the whole sample, measured on a 0–100 scale where higher values indicate perceptions that LLMs create more benefits than harms. Panel B shows estimated effects

**from a linear regression model predicting attitudes toward LLMs from regular use, age, gender, education level, and political orientation. Points represent the regression coefficients along with 95% confidence intervals. Opaque points indicate statistically significant effects (p < .05), while faded points indicate non-significant effects.**

### 3.7.1 SOCIAL AND DEMOGRAPHIC PREDICTORS OF ATTITUDES TO LLMS

We examined whether regular LLM use, age, gender, education level, and political orientation were associated with attitudes to LLMs across the whole sample using a linear regression model.

Regular users of LLMs were significantly more positive about LLMs than non-regular users, scoring on average nearly 19 points higher on the attitudes scale ($b = 18.90$, $SE = 1.23$, $p < .001$). Attitudes were also slightly more positive among older respondents, with each additional decade of age associated with a 1.65 point increase on the scale (though this effect size is small) ($b = 1.65$, $SE = 0.37$, $p < .001$). Male participants also reported more positive attitudes, ($b = 7.35$, $SE = 1.11$, $p < .001$), as did more politically right-leaning respondents ($b = 1.07$, $SE = 0.25$, $p < .001$). The effect of education was not significant ($b = 0.18$, $SE = 1.13$, $p = .876$).

The above results are corroborated by two additional logistic regression models examining the association between the same predictor variables and responses relating specifically to perceptions of the benefits and risks of LLMs. In the model examining predictors of the perceived benefits of LLMs, regular users ($OR = 6.82$, $p < .001$), male respondents ($OR = 1.28$, $p = .045$) and politically more right-leaning respondents ($OR = 1.12$ per 10-point increase on the political scale, $p < .001$) were significantly more likely to report enthusiasm for the potential benefits of LLMs, while effects of age and education were not significant.

In the model predicting LLM concern, regular users were significantly less likely to report being concerned about the potential risks of LLMs ($OR = 0.35$, $p < .001$), as were older respondents ($OR = 0.88$ per decade increase in age, $p < .001$), male respondents ($OR = 0.71$, $p < .001$), and more right-leaning respondents ($OR = 0.92$ per 10-point increase on the political scale, $p < .001$), while the effect of education was not significant. Figure 4, panel B, shows demographic predictors of attitudes to LLMs. Full results from these regressions are presented in Table 3 of the Supplementary Information.

# 4. Discussion

Using a nationally representative survey of UK adults, we examined public adoption of LLMs for both practical and social purposes, public trust in the accuracy of information generated by LLMs, and overall attitudes towards these technologies in terms of their perceived societal benefits and risks. Across each of these areas, we also examined the role of age, gender, educational attainment, and political orientation in shaping patterns of use, trust, and sentiment.

## 4.1 PUBLIC ADOPTION OF LLMS

We found that a large majority of respondents had used an LLM at least once before. Among these, use was typically frequent, with almost three quarters (73.4%) reporting at least weekly use. For regular users, LLMs were most commonly used for practical purposes, such as work-related tasks, information seeking, planning activities, and brainstorming ideas. However, interacting with LLMs for social and emotional purposes was not uncommon. Almost one third of regular users (31%) reported turning to LLMs for personal or emotional support at least occasionally, while one quarter reported using them for meaningful conversations. Fewer users engaged with LLMs specifically for roleplay, although a notable minority reported doing so at least occasionally (14%). When asked how human-like interactions with LLMs typically feel, more than eight in ten regular users described them as at least slightly human-like, with many reporting that they felt moderately human-like.

Age was the most consistent predictor of LLM use, with the likelihood of engagement decreasing with age across all five types of use. Men were significantly more likely to use LLMs as practical tools and for roleplay than women, but less likely to use them for personal support. Respondents with a university degree were more likely to use LLMs for practical purposes than those without a degree.

While some prior work examines the extent of LLM use across multiple countries (e.g., [2, 1, 16]), few studies have compared practical forms of use with more social or emotional ones. We find that while practical activities remain far more common, substantial numbers of people are turning to LLMs for personal support, companionship, and meaningful conversation. This is important because amid growing concerns about the potential for these tools to alter patterns of social connection and human relationships [12], our findings suggest that this type of use is already commonplace among many users. While relatively little is known about the longer-term consequences of using LLMs for social purposes, concerns have been raised that social forms of AI interaction may encourage emotional reliance on AI systems or adversely shape users' expectations of human relationships [28, 29, 30]. These findings highlight the need for further research into the potential benefits and risks associated with social and emotional engagement with LLMs, including when such interactions may complement existing sources of support and when they may risk displacing engagement with other people or professional services.

## 4.2 PUBLIC TRUST IN LLMS

We found relatively high levels of trust in information generated by LLMs across the topics examined. For information and advice relating to elections, public services, climate change, legal matters, and health issues, majorities of respondents reported trusting LLMs at least somewhat (63%–70%). Trust was on the whole lower for information and advice relating to personal relationships, although 38% of respondents reported at least some trust in this area, and almost one in ten reported high trust. Compared with other sources, LLMs were generally trusted less than experts, online search, and friends and family across topics (with the exception of climate change, where trust in LLMs exceeded trust in friends and family), but more than newspapers and television, and substantially more than social media. LLM use was the strongest predictor of trust in LLM-generated information, with regular users significantly more likely to trust LLM-generated information and advice across all contexts. For some topics, trust also increased with age, with older adults reporting greater trust in LLMs for information about elections, legal matters, and health issues, though these effects were considerably smaller.

While several studies suggest that seeking information is now one of the most common uses of LLMs among the public across multiple countries [18, 16], relatively little is known about the extent to which people trust the information and advice these systems provide. Our findings suggest that people already trust LLM-generated information across a range of domains, including those that carry personal consequences such as health and legal matters. This is noteworthy because if trust shapes the extent to which people rely on and act upon given information, LLM use may increasingly influence human behaviour across a range of contexts. Because it is well-documented that LLMs are capable of generating inaccurate or misleading outputs [31, 20], it is crucial to ensure that users are aware of the limitations of these systems and exercise appropriate caution.

It will be important for further research to examine not only trust in different sources of information, but also when and why people choose to consult them. While trust is likely to be one factor shaping information-seeking behaviour, other considerations such as accessibility, availability, speed, and ease of interaction may also play important roles. As such, a readily available source that is trusted enough may in practice be chosen over a more trusted source that is more difficult to access. Our findings offer a foundation for understanding how people balance trust, convenience, and perceived expertise when seeking information in the age of generative AI.

## 4.3 PUBLIC ATTITUDES TOWARDS LLMS

Overall, attitudes towards LLMs were somewhat more positive than negative, although views varied considerably across respondents. While over three quarters of the sample (77.3%) reported feeling enthusiastic about the potential benefits of LLMs, a high proportion (69.6%) also reported feeling concerned about their potential risks. Our findings suggest that many people do not hold simply positive or negative views about LLMs and their societal impacts, but instead recognise both their opportunities and their risks. This pattern is consistent with previous research on public attitudes towards AI more generally [3, 22, 2], which often finds that people express both optimism and concern about the role of AI in society.

It is noteworthy that, after the extent of LLM use itself, gender emerged as the strongest and most consistent predictor of attitudes towards these technologies. Male respondents were more likely to report feeling enthusiastic about the potential benefits of LLMs, less likely to express concern about their potential risks, and more likely to believe that LLMs will generate more benefits than harms overall. These findings are consistent with previous research suggesting that women often perceive greater risk in digital environments more broadly [24, 25, 32], with much work suggesting this increased caution reflects the fact that technology-facilitated harms disproportionately affect women, reflecting and reproducing existing structural inequalities within digital spaces [26, 33, 34]. The presence of significant gender differences in attitudes towards LLMs has important implications for digital equity and inclusion. If generative AI is to deliver equal benefits across society, it will be important to ensure that all groups feel able to safely engage with these technologies.

Our findings offer an up-to-date overview of public sentiment towards generative AI in the UK, showing that while many people are enthusiastic about the potential benefits of these technologies, concerns about their risks remain widespread. Future research would benefit from examining public expectations for the governance and regulation of generative AI in greater detail, including the roles that governments, regulators, and technology companies should play in ensuring that these systems are safe [35].

## 4.4 STUDY LIMITATIONS

While our work offers novel insights into public engagement with LLMs in the UK, it is important to acknowledge the limitations of the research, as well as outstanding questions.

Our Prolific sample was designed to be broadly representative of the UK population in terms of age, gender, and region. However, Prolific recruitment is not probability-based and instead relies on participants voluntarily joining the platform and completing studies on a first-come, first-served basis. Participants are therefore likely to be more digitally engaged than the general population, with regular internet access and sufficient digital literacy to navigate online platforms. As a result, respondents in our sample may spend more time online and be more familiar with generative AI tools than the general population. This is reflected in our estimates of LLM adoption, which are higher than those reported in other recent studies (e.g., [3, 2]). As a result, self-reported overall use rates presented in our findings are unlikely to be representative of the UK population as a whole. Rather, our results provide insight into how people who use LLMs typically engage with these tools in practice.

Additionally, our sample included only adults. Given growing concerns about the implications of LLM use for learning, education, social development, and mental wellbeing among children and adolescents [36, 37, 38], it will be important for future research to examine how younger populations engage with these technologies. In particular, work is needed to understand the types of interactions children and teenagers typically have with LLMs, the extent to which they already trust and rely upon these systems, and whether and how LLM use influences outcomes relating to cognitive, social and emotional development. As younger generations are now likely to grow up with generative AI as a routine part of everyday life, understanding these issues represents an important priority for future research.

Finally, as already noted, this is an area of study that is rapidly evolving, meaning that key findings may change as adoption continues to increase at both individual and organisational levels, and with developments in media and policy discourse surrounding these technologies. The findings presented here therefore capture public engagement with LLMs at a particular moment in time, and we highlight the importance of regular monitoring as patterns of use, trust, and sentiment continue to change.

# 5. Conclusion

Taken together, our findings show that LLMs are already used across many aspects of life, with practical applications remaining the most common form of activity, but with substantial numbers of users also turning to LLMs for emotional support, meaningful conversation, and social companionship. We also find relatively high levels of trust in information generated by LLMs across a range of domains, alongside attitudes characterised by both optimism about the potential benefits of these technologies and concern about their possible risks. Finally, we identify some demographic differences in patterns of use, trust, and sentiment, highlighting that experiences of generative AI are unlikely to be uniform across society.

At a time when the societal impacts of generative AI are the subject of growing public, academic, and policy attention, these results provide important new evidence on how people in the UK are currently engaging with LLMs, while also offering a baseline against which changes in use, trust, and public sentiment can be assessed. The findings particularly highlight the need for a greater understanding of the consequences associated with long-term LLM use, not only for learning and cognition, but also for social relationships, given that social and emotional engagement with LLMs is already relatively common. Additionally, the findings point to the need for further research on how trust in these systems shapes the extent to which people rely on and act upon the information and advice they provide compared to other sources. Finally, the presence of demographic differences relating to age, gender, and educational attainment in patterns of use, trust, and sentiment shows the importance of working to ensure that these tools are accessible and beneficial to all members of society.

# Acknowledgements

This work was supported by the AI Security Institute and the Ecosystem Leadership Award under the EPSRC Grant EPX03870X1 at the Alan Turing Institute. We are grateful to Professor Christopher Summerfield and Tvesha Sippy for valuable discussions on this work.

# Supplementary Information

## A1 DESCRIPTION OF LLMS GIVEN TO PARTICIPANTS

At the start of the survey, participants were provided with the following description of LLMs: *AI chatbots (technically called large language models, or LLMs) are computer programs that use artificial intelligence to generate text, speech, or images in response to your questions or prompts. They can hold extended conversations, answer questions, and help with tasks. These tools may appear as standalone chatbots or be built into websites and apps. Examples include ChatGPT, Google Gemini, Claude, Microsoft Copilot, Perplexity, Grok, and those found on platforms like Character.ai, and Replika, among others.*

*Please note, we are not referring to voice assistants like Alexa and Siri which do not generate content in the same way as the above examples and do not use the same underlying language model technology.*

Participants were also provided with the following description of AI companions in the follow-up question:

*More specifically, have you ever used an AI chatbot in the form of an AI companion (a chatbot specifically designed to provide emotional support, friendship and conversation) on a character-based platform (like Character.AI, Replika or similar)?*

## A2 TABLES SHOWING FULL REGRESSION RESULTS

Table 1 presents the full regression results for the demographic predictors of types of LLM use discussed in Section 3.5.1 of the main report.

Table 2 presents the full regression results for the demographic predictors of trust in information given by LLMs, noted in Section 3.6.1 of the main report.

Table 3 shows full regression results for the demographic predictors of overall attitudes to LLMs, reported in Section 3.7.1 of the main report.

**Table 1:** Logistic regression models examining demographic predictors of types of LLM interaction amongst regular users.

| | Practical tasks | | Collaboration | | Support | | Conversation | | Roleplay | |
|---|---|---|---|---|---|---|---|---|---|---|
| Predictor | OR (95% CI) | p | OR (95% CI) | p | OR (95% CI) | p | OR (95% CI) | p | OR (95% CI) | p |
| Age | 0.78 (0.68–0.89) | < .001*** | 0.82 (0.76–0.90) | < .001*** | 0.74 (0.68–0.80) | < .001*** | 0.84 (0.77–0.92) | < .001*** | 0.77 (0.69–0.86) | < .001*** |
| Gender (Male) | 1.48 (0.99–2.23) | .060 | 0.91 (0.70–1.18) | .490 | 0.68 (0.53–0.86) | .002** | 1.13 (0.87–1.46) | .361 | 1.51 (1.09–2.10) | .014* |
| Education (Degree) | 1.86 (1.24–2.80) | .003** | 1.45 (1.11–1.88) | .006** | 1.10 (0.86–1.42) | .460 | 1.12 (0.86–1.47) | .386 | 1.31 (0.94–1.86) | .119 |
| Political orientation | 0.99 (0.90–1.09) | .868 | 1.05 (0.99–1.12) | .094 | 0.99 (0.94–1.05) | .726 | 1.04 (0.98–1.11) | .160 | 1.09 (1.01–1.17) | .027* |
| N | 1307 | | 1304 | | 13.09 | | 1303 | | 1306 | |
| Pseudo $R^2$ | 0.04 | | 0.02 | | 0.04 | | 0.01 | | 0.03 | |

**Note.** Odds ratios (OR) from logistic regression models examining demographic predictors of different types of LLM interaction amongst regular users of LLMs. The outcome in each model indicates whether respondents reported using LLMs in that way at least occasionally (very often or occasionally vs. very rarely or never). Age is scaled per 10-year increase and political orientation is scaled per 10-point increase on a continuous left–right scale. Gender is coded as male (reference = non-male) and education as holding a university degree (reference = no degree). Pseudo $R^2$ values are McFadden's pseudo $R^2$. *$p < .05$, **$p < .01$, ***$p < .001$.

**Table 2:** Logistic regression models examining demographic predictors of trust in information provided by LLMs across six information domains.

| | Public Services | | Elections | | Climate | | Legal | | Health | | Relationships | |
|---|---|---|---|---|---|---|---|---|---|---|---|---|
| **Predictor** | **OR (95% CI)** | **p** | **OR (95% CI)** | **p** | **OR (95% CI)** | **p** | **OR (95% CI)** | **p** | **OR (95% CI)** | **p** | **OR (95% CI)** | **p** |
| LLM use (Regular) | 3.39 (2.29–5.06) | < .001*** | 3.27 (2.26–4.76) | < .001*** | 4.17 (2.84–6.16) | < .001*** | 4.54 (3.10–6.71) | < .001*** | 3.64 (2.51–5.34) | < .001*** | 3.52 (2.35–5.36) | < .001*** |
| Age | 1.10 (0.97–1.24) | .132 | 1.12 (1.00–1.26) | .042* | 1.08 (0.96–1.22) | .220 | 1.13 (1.00–1.28) | .045* | 1.14 (1.02–1.28) | .023* | 0.91 (0.81–1.03) | .128 |
| Gender (Male) | 1.49 (1.03–2.15) | .033* | 0.93 (0.66–1.31) | .678 | 0.70 (0.49–1.01) | .055 | 1.20 (0.83–1.73) | .339 | 1.28 (0.90–1.81) | .173 | 0.98 (0.70–1.37) | .893 |
| Education (Degree) | 0.83 (0.57–1.21) | .342 | 0.76 (0.53–1.08) | .125 | 0.95 (0.66–1.37) | .790 | 0.88 (0.61–1.28) | .509 | 0.72 (0.50–1.03) | .074 | 1.21 (0.86–1.72) | .279 |
| Political orientation | 1.08 (0.99–1.18) | .071 | 1.12 (1.04–1.21) | .005** | 1.02 (0.94–1.10) | .657 | 1.12 (1.03–1.22) | .011* | 1.06 (0.98–1.15) | .135 | 1.11 (1.02–1.20) | .012* |
| N | 637 | | 648 | | 647 | | 638 | | 643 | | 643 | |
| Pseudo $R^2$ | 0.07 | | 0.07 | | 0.07 | | 0.10 | | 0.07 | | 0.07 | |

**Note.** Odds ratios (OR) from logistic regression models examining demographic predictors of trust in information provided by LLMs across six information domains. The outcome in each model indicates whether respondents reported trusting information provided by LLMs at least somewhat (very much or somewhat vs. not at all or not very much). Age is scaled per 10-year increase and political orientation per 10-point increase on a continuous left–right scale. Gender is coded as male (reference = non-male) and education as holding a university degree (reference = no degree). Pseudo $R^2$ values are McFadden's pseudo $R^2$. $^*p < .05$, $^{**}p < .01$, $^{***}p < .001$.

**Table 3:** Regression models examining demographic predictors of attitudes towards LLMs.

| | Attitudes (overall, continuous) | | Perception of Benefit | | Perception of Risk | |
|---|---|---|---|---|---|---|
| Predictor | b (SE) | p | OR (95% CI) | p | OR (95% CI) | p |
| LLM use (Regular) | 18.90 (1.23) | < .001*** | 6.82 (5.34–8.76) | < .001*** | 0.35 (0.27–0.45) | < .001*** |
| Age | 1.65 (0.37) | < .001*** | 0.97 (0.89–1.05) | .391 | 0.88 (0.83–0.95) | < .001*** |
| Gender (Male) | 7.35 (1.11) | < .001*** | 1.28 (1.01–1.63) | .045* | 0.71 (0.58–0.87) | < .001*** |
| Education (Degree) | 0.18 (1.13) | .876 | 0.96 (0.75–1.22) | .746 | 1.17 (0.95–1.44) | .143 |
| Politicical orientation | 1.07 (0.25) | < .001*** | 1.12 (1.06–1.18) | < .001*** | 0.92 (0.88–0.96) | < .001*** |
| N | 1903 | | 1924 | | 1918 | |
| Model fit ($R^2$ / Pseudo $R^2$) | 0.16 | | 0.15 | | 0.05 | |

**Note.** Regression models examining demographic predictors of attitudes towards LLMs. The first column presents coefficients from a linear regression predicting overall attitudes toward LLMs (0–100 scale). The second and third columns present odds ratios from logistic regression models predicting the extent to which respondents were enthusiastic about the potential benefits of LLMs or were concerned about the potential risks of LLMs (for both: strongly agree/somewhat agree vs. strongly disagree /somewhat disagree). Age is scaled per 10-year increase and political orientation per 10-point increase on a continuous left–right scale. Gender is coded as male (reference = non-male) and education as holding a university degree (reference = no degree). $^{*}p < .05$, $^{**}p < .01$, $^{***}p < .001$.